\documentstyle[epsfig,aps,amsmath,twocolumn,floats]{revtex}

\begin{document}
\draft
\wideabs{
\title{Drying Induced Hydrophobic Polymer Collapse}
\author{Pieter Rein ten Wolde and David Chandler}
\address{Department of Chemistry, University of California,
Berkeley,\\
California 94720}
\maketitle

\begin{abstract}
We have used computer simulation to study the collapse of a
hydrophobic chain in water. We find that the mechanism of collapse is
much like that of a first order phase transition. The evaporation of
water in the vicinity of the polymer provides the driving force for
collapse, and the rate limiting step is the nucleation of a
sufficiently
large vapor bubble. The study is made possible through the
application of
transition path sampling and a coarse grained treatment of liquid
water.
Relevance of our findings to understanding the folding and assembly of
proteins is discussed.
\end{abstract}


}
\newpage

For nearly a half-century, hydrophobic interactions have been
considered the
primary cause for self assembly in soft matter, and a major source of
stability in biophysical assembly~\cite{Kauzmann58,Tanford}. Studying
these
interactions in perhaps their most basic form, we use computer
simulations
to demonstrate the mechanism for the collapse of a hydrophobic
polymer in
water. We show that solvent fluctuations induce the transition from
the
extended coil to the collapsed globule state, where a vapor bubble of
sufficient size is formed to envelop and thereby stabilize a critical
nucleus of hydrophobic units. This mechanism is different from that
usually
considered, where coil to globule transitions are attributed to
effective
interactions between pairs of chain segments and a change in sign of
second
virial coefficient~\cite{Grosberg94}. Rather, the mechanism we find is
evocative of the n-cluster model, where hydrophobic collapse is
produced by
solvent induced interactions between a relatively large cluster of
segments~\cite{DeGennes91}.

As expected from earlier work on the equilibrium theory of
hydrophobicity~\cite{Lum99}, we find that the solvent length scales pertinent to
hydrophobic collapse extend over nanometers. We also find that
pertinent
time scales extend beyond nanoseconds. Given these molecularly large
lengths and times,
it is understandable that no work before this has provided
statistically
meaningful computer simulations of the process. Our use of a
statistical
field model of water allows us simulate solvent dynamics over large
length
and time scales that would be impractical to study with purely
atomistic
simulation. Spatially complex small length scale fluctuations are
analytically integrated out,
thus removing the most computationally costly features from our
simulation.
Their
integration can be performed at the outset because their relaxation is
relatively fast~\cite{Fast} and their statistics is essentially
Gaussian~\cite{Hummer96}. Only the polymer degrees of freedom and a
coarse-grained
density field remain. The equilibrium theory for this approach has
been
detailed in an earlier paper~\cite{TenWolde01}. Here, we use a
version of
the model that is suitably generalized for dynamical applications.

By ``small length'' we refer to distances smaller than $l\approx $
0.3~nm. In
the absence of any strong perturbation, such as those that can occur
close
to a solute, these small length scale fluctuations are the only
fluctuations
of significance. Larger length scale fluctuations are generally
insignificant in water at ambient conditions. The liquid is
relatively cold
and incompressible, and spatial correlations extend over only one or
two
water molecules. But the presence of a hydrophobic polymer can change
this
situation, making large length scale fluctuations important. The
liquid lies
close to macroscopic vapor-liquid equilibrium, and vapor-like
behavior is
stabilized in the vicinity of a sufficiently extended hydrophobic
surface~\cite{Stillinger}, a surface formed, for example, by the clustering of
hydrophobic groups in a polymer chain. This effect can be captured by
the
behavior of a coarse-grained density field~\cite{Lum99,Weeks98}, and
such a
field is conveniently simulated with a field of binary numbers~\cite
{TenWolde01}, as we discuss now.

\section*{Equilibrium of the solvent model}

Dividing space into a cubic grid of cells, each with side length $l$,
this binary field is a dynamical variable given by
$\rho _{l}n_{i}$, where $\rho _{l}$ is a constant given by
the bulk liquid density, and $n_{i}$ is dynamical given by $n_{i}=1$
if cell
$i$ contains ``liquid,'' and $n_{i}=0$ if it contains ``vapor''. This
coarse-grained density is thus the ``Ising'' field that describes
phase
coexistence and gas-liquid interfaces\cite{Chandler87}. The remaining
small
length scale field, $\delta \rho ({\bf r})$, is the difference
between $\rho
_{l}n_{i}$ and the actual density at a point ${\bf r}$ in cell $i$.
It takes
on a continuum of values, obeys Gaussian statistics, and captures the
granularity of the solvent~\cite{Chandler93}.

Confining our attention to an ideal hydrophobic solute, namely a
polymer
consisting of hard spheres, the integration over $\delta \rho \left(
{\bf r}\right) $ yields the following Hamiltonian for the coarse-grained
field in
the grand-canonical ensemble~\cite{TenWolde01}:
\begin{equation}
H\left[ \left\{ n_{k}\right\} ;\left\{ v_{i}\right\} \right] \approx
\sum_{i}\left[ -\mu +\Delta \mu _{\mbox {\tiny ex}}(v{_{i}})\right]
\,n_{i}-\epsilon \sum_{<i,j>}n_{i}n_{j}.  \label{eq:H_dmu}
\end{equation}
The quantity $\mu $ is the solvent chemical potential, $\epsilon $ is
the
nearest-neighbor coupling energy parameter of the solvent, the sum
over $\left\langle i,j\right\rangle $ includes all nearest neighbor cells,
$v_{i}$
denotes the volume excluded by the solute in cell $i$, and
\begin{equation}
\Delta \mu _{\mbox {\tiny ex}}(v{_{i}})\approx cv_{i},
\label{eq:deltamu}
\end{equation}
with $c=2.67\times 10^{8}~$J/m$^{3},$ is the reversible work to
accommodate
that excluded volume in the liquid. Equations (\ref{eq:H_dmu}) and
(\ref
{eq:deltamu}) are approximations to the full results derived in
Ref.~\cite
{TenWolde01}. Specifically, in comparison with Eqs.(19) or (23) of
Ref.~\cite
{TenWolde01}, we see that Eq.(\ref{eq:H_dmu}) neglects relatively
small
inter-cell interactions. Further, Eq.(\ref{eq:deltamu}) is a
reasonable
numerical approximation to the general result for $\Delta \mu _{\mbox {\tiny
ex}}(v{_{i}})$ that follows from Gaussian statistics for $\delta \rho
\left(
{\bf r}\right) $~\cite{Hummer96}. The specific value of $c$ is chosen
so
that $\Delta \mu _{\mbox {\tiny ex}}(4\pi R^{3}/3)$ yields the excess
chemical potential for a hard sphere or bubble that excludes solvent
from a
spherical volume of radius $R=$0.5 nm~\cite{Huang01_1}.

\begin{figure}[t]
\begin{center}
\epsfig{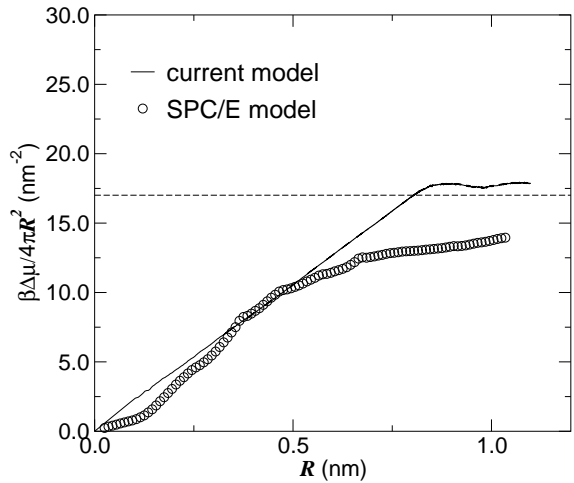}
\end{center}
\caption{\label{fig:bdmuoA_R.IS} The solvation energy per unit surface area for a hard sphere
or
bubble as a function of $R$, where $R$ is the radius of the spherical
volume
from which the solvent, water, is excluded. The circles are the
results of
computer simulation for the SPC/E atomistic model of water {\protect\cite
{Huang01_1}}. The line is the prediction of our model computed by
Monte Carlo umbrella sampling~\protect\cite{Frenkel02}, using the weight
functional
$\exp \left\{ -\beta H\left[ \left\{ n_{k}\right\} ,\left\{
v_{i}\right\}\right] \right\} ,$where $\beta ^{-1}=k_{B}T$, and with $\Sigma
_{i}v_{i}=4\pi R^{3}/3$.The parameters of the model have been
chosen so as to match the isothermal compressibility, the chemical
potential and the surface tension of water at room temperature and 1
atm pressure. This yields $l=0.21$nm, $\epsilon =3.74$kJ/mol
($=1.51k_{B}T$ at room temperature), and $\mu -\mu _{{\rm
coex}}=0.557$J/mol ($=2.25\times 10^{-4}k_{B}T$ at room temperature),
where $\mu _{{\rm coex}}$ is the chemical potential of the solvent
when it is in coexistence with the vapor at room temperature. The
comparison of $\mu -\mu _{{\rm coex}}$ with $\epsilon $ of shows that
the bulk liquid is extremely close to phase coexistence. The
horizontal dashed line is placed at the value of the liquid-vapor
surface tension and coincides with the large $R$ asymptotic value of
$\Delta \mu _{{\rm ex}}/4\pi R^{2}$ for both the model and the
atomistic simulation (the surface tension of SPC/E water model is
very close to that of real water\protect\cite{Alejandre95}).}
\end{figure}

The predictions of the model for excess chemical potential or
solvation free
energy of a hard sphere over the range of all possible radii is shown
in
Fig.~\ref{fig:bdmuoA_R.IS}. It is seen that the agreement between the
results of the model and those of an atomistic simulation is
reasonable, as
the model captures the appropriate free energy scaling at small and
large
lengths and gives a reasonable prediction of the crossover length
between
the two regimes. For closer agreement between the statistical field
model
and the atomistic model in the crossover regime, terms neglected in
Eq.(\ref
{eq:H_dmu}) should be included, as demonstrated in
Ref.~\cite{TenWolde01}.

The crossover occurring around $R\sim 1~$nm is a collective effect.
It is
induced by the entropic cost of constraining small length scale
fluctuations
near a hydrophobic solute. For a given cell $i,$ the cost is $\Delta
\mu _{\mbox {\tiny ex}}(v{_{i}})$. When the $v_{i}$'s form a connected
volume of
large enough extent, the net entropic cost is larger than the
energetic cost
for forming an interface that surrounds the volume. At this point,
and for
larger volumes, it becomes energetically preferable for $n_{i}=0$ for
all
cells $i$ occupied by or adjacent to the large solute, and the free
energetics is then dominated by the energy of the interface thus
formed.
Such interfaces are stable only in the neighborhood of vapor-liquid
phase equilibrium.
Therefore, crossover behavior can occur only when $\mu -\mu _{{\rm
coex}}\ll
k_{B}T$, where $k_{B}T$ is Boltzmann's constant times temperature
(the thermal
energy) and $\mu _{{\rm coex}}$ is the chemical potential at phase
coexistence between liquid and vapor. Further, since a
pressure-volume term
generally contributes to the free energy of a large void, crossover
behavior
occurs only when pressure is low enough to make this term similarly
small.
We will see below that the crossover in the scaling behavior of the
free
energy is closely related to the nucleation of oil-water demixing and
the
mechanism for the collapse of a hydrophobic chain.

\section*{Dynamics of the solvent and chain}

The excluded volumes $v_{i}$ are dynamical, changing as the chain
configuration changes with time. The chain need not move slowly on
the time scale for which disturbances in the solvent may relax. As such,
reversible work surfaces, like the solvation free energy graphed in Fig.~\ref
{fig:bdmuoA_R.IS}, are not entirely pertinent, and the dynamics of the
liquid and chain should be studied together. For this purpose, we have
constructed a stochastic dynamics that satisfies the basic
requirements of time reversibility and the preservation of an equilibrium
distribution. With this dynamics, solute particles move through space 
with continuous
variation of coordinates while the coarse grained density evolves
simultaneously through its discrete configurations. Since $\delta \rho
\left( {\bf r}\right) $ has been integrated out (its effects appear
only
implicitly) the trajectories can be true to nature only for times
greater
than those required to relax $\delta \rho \left( {\bf r}\right) $.
This time
is of the order of $1$~ps~\cite{Fast}. On this time scale, the
appropriate
stochastic dynamics is non-inertial.

Trajectories of our model are advanced as follows: The polymer,
composed of $N_{{\rm s}}=12$ spheres is moved for $M_{{\rm s}}$ time steps from a
configuration $\left\{ {\bf r}_{\alpha }\right\} $ to a new
configuration $\{{\bf r}_{\alpha }^{\prime }\}$ according to propagation rules of
Langevin
dynamics, in the field of constant coarse grained density variables,
$\left\{ n_{i}\right\} $. At the end of those steps, the polymer is
held
fixed and the coarse grained density is moved to a new configuration
$\left\{ n_{i}^{\prime }\right\} $ by carrying out a full sweep
through all
cells, applying Glauber Monte Carlo dynamics~\cite{Binder88} to each
cell
with the Hamiltonian $H\left[ \left\{ n_{i}\right\} ,\left\{
v_{i}\right\}
\right] $. The net procedure taking the system from $\left\{ {\bf
r}_{\alpha
},n_{i}\right\} $ to $\left\{ {\bf r}_{\alpha }^{\prime
},n_{i}^{\prime
}\right\} $ is then repeated over and over again. The trajectories
thus
formed are reversible, preserve the norm of the configurational
distribution
function, and obey detailed balance.

For one of the $M_{{\rm s}}$
 time steps between solvent sweeps,
carrying the
monomers from their configuration at time $t$ to that at time
$t+\delta t_{
{\rm s}}$, the Langevin propagation corresponds to
\begin{equation}
{\bf r}_{\alpha }(t+\delta t_{{\rm s}})={\bf r}_{\alpha }(t)+\delta
t_{{\rm s}}\gamma ^{-1}{\bf F}_{\alpha }(\left\{ {\bf r}_{\xi }(t)\right\}
,\left\{
n_{i}\right\} ),  \label{eq:langevin}
\end{equation}
where ${\bf F}_{\alpha }(\left\{ {\bf r}_{\xi }(t)\right\} ,\left\{
n_{i}\right\} )$ is the force acting on monomer $\alpha $, and
$\gamma $ is
the friction constant. The force contains a random part, $\delta {\bf
F}$,
that is independent of configurational variables. It is the dynamical
remnant of the small length scale field, $\delta \rho \left( {\bf
r}\right) $.
We take it to be isotropic, to have a correlation time of zero, and
to obey
Gaussian statistics with zero mean and variance $\left\langle \left|
\delta
{\bf F}\right| ^{2}\right\rangle =3k_{B}T\gamma $. We estimate the
value of $\gamma $ by associating the diffusion constant of a single bead
moving in
water, $D_{{\rm s}}=k_{B}T/\gamma $, to its Stokes estimate which
gives $\gamma =8.39\times 10^{-12}$kg/s. The remaining contributions to the
force
is not random, and is given by $-\nabla _{\alpha }\left[ V\left(
\left\{
{\bf r}_{\xi }\right\} \right) +c\sum_{i}v_{i} n_i\right] $. The time
step is $\delta t_{{\rm s}}=1.4\times 10^{-14}$~s, which is small enough that
(\ref
{eq:langevin}) preserves the canonical distribution.

A physically meaningful value of $M_{{\rm s}}$ coincides with
$M_{s}\approx
\delta t_{l}/\delta t_{{\rm s}}$, where $\delta t_{l}$ is the
physical time
associated with a sweep through the coarse grained density of the
solvent.
This solvent time should coincide with the correlation time for a
density
fluctuation of length scale $l$, i.e, $\delta t_{l}=1/[D(2\pi
/l)^{2}],$
where $D$ is the self diffusion constant of liquid water. This
relationship
gives $\delta t_{l}=5.0\times 10^{-13}$s, and thus $M_{{\rm s}}=36$.

With these dynamical rules, we show below that the half-life of the
extended chain we consider is approximately 10$^{-5}~$s at room
temperature, while the
transient
time to move off a threshold between extended coil and compact globule
states is of the order of 10$^{-10}$~s. Thus, collapse transitions in
this
system are rare events\cite{Chandler78}. Harvesting a representative
ensemble of these events without prior knowledge of transition states
or
mechanism is accomplished with transition path
sampling~\cite{Bolhuis01_1}.
To apply this technique, we use the number of vapor cells, $\Sigma %
_{i}(1-n_{i})$, as the order parameter to distinguish the solvated
coil and
globule states.

\section*{Results}

\begin{figure}[h]
\begin{center}
\epsfig{figure = 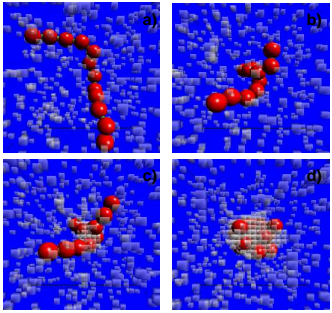,width=4.35cm}
\end{center}
\caption{Four configurations from a
trajectory where a 12 unit hydrophobic chain in water goes from the
coil to the globule state. (a) shows a configuration from the
equilibrated coil. The chain remained in configurations like that
throughout a $10~$ns run at room temperature ($%
T=0.663\epsilon $).
On a much longer time scale, about $10^{-5}~$s, the chain typically
does exhibit a transition from the coil to globule. Such events are
found with transition path sampling, equilibrating from an initial
high temperature ($T=0.74\epsilon )$ 10~ns trajectory that exhibited
the transition spontaneously. Three configurations exhibiting that
transition, covering a stretch of $1.5~$ns are shown in (b), (c) and
(d),
with that in (c) being a configuration from the transition state
surface.
The transparent cubes denote the vapor cells. Those seen far from the
chain
are typical spontaneous density fluctuations in bulk liquid water.
The size
of the simulation box is $397~{\rm nm}^{3}$, corresponding to 42,875
cells.
The parameters characterizing the energy of the solvent are given in
the
caption to Fig.1. The intra-chain potential, $V\left( {\bf
r}_{1},{\bf r}%
_{2},...,{\bf r}_{12}\right) $, is a function of the positions of the
centers of each of the 12 hydrophobic spheres (the red particles in
the
figure). It contains three parts: (1) steep (essentially hard sphere)
repulsions between solute particles such that their interparticle
separations are larger than $\sigma =0.72~$nm; (2) stiff harmonic
potentials
bonding adjacent particles in the hydrophobic chain, $\frac{1}{2}%
k_{s}(\sigma -\left| {\bf r}_{\alpha +1}-{\bf r}_{\alpha }\right|
)^{2}$,
with $k_{s}=14.1$J/m$^{2}$; (3) a bending potential favoring an
extended
chain, $\frac{1}{2}k_{\theta }\theta _{\alpha }{}^{2},$ where $\theta
_{\alpha }$ is the angle between $\left( {\bf r}_{\alpha +2}-{\bf
r}_{\alpha
+1}\right) $ and $\left( {\bf r}_{\alpha +1}-{\bf r}_{\alpha }\right)
$, and
$k_{\theta }=1.85\times 10^{-20}$J/rad$^{2}$. The volumes $v_{i}$
excluded
from water by the chain are dynamic as they change with changing chain
configuration, i.e., $v_{i}=v_{i}\left( \left\{ {\bf r}_{\alpha
}\right\}
\right) $. Specifically, these volumes are computed by assuming water
molecules have van der Waals radii equal to $0.14$nm, and that the
diameter
of each hydrophobic unit is $\sigma =0.72~$nm.. That is, points in the
excluded volume, ${\bf r}$, are those in the union of all volumes
inscribed
by $|{\bf r}-{\bf r}_{\alpha }|<0.5~$nm, $\alpha =1,2,...,N_{{\rm
s}}$.}
\label{fig:tps.path}
\end{figure}

\begin{figure}[t]
\begin{center}
\epsfig{figure = 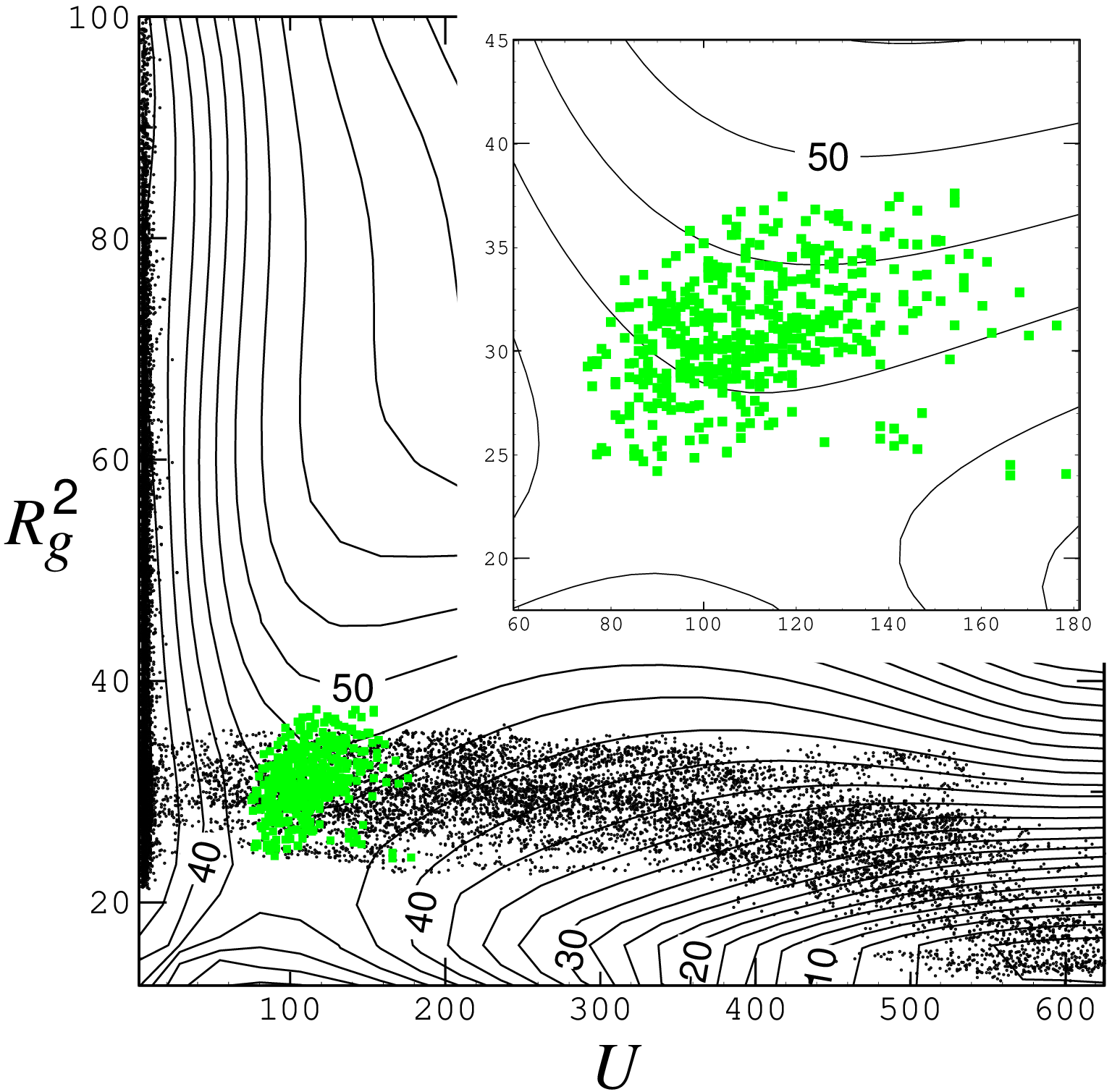,width=5cm}
\end{center}
\caption{\label{fig:umbr_tps}Contour plot of the free energy landscape
for the collapse of our hydrophobic polymer, computed by Monte Carlo
umbrella sampling~\protect\cite {Frenkel02}, using the weight
functional $\exp \left\{ -\beta H\left[ \left\{n_{k}\right\} ,\left\{
v_{i}\right\} \right] -\beta V\left[ \left\{ {\bf r}_{\alpha }\right\}
\right] \right\} $, where the intrachain potential, $ V\left[ \left\{
{\bf r}_{\alpha }\right\} \right] $, and the excluded volumes,
$v_{i}\left[ \left\{ {\bf r}_{\alpha }\right\} \right] $, are
determined as described in the legend to Fig. 2. The curves of
constant free energy are drawn as a function of the squared radius of
gyration and the size of the largest bubble in the system (see
text). Neighboring lines are separated by $2.5k_{B}T$. Superimposed is
a scatter plot (in black) of the harvested 150~ps trajectories going
from the coil to the globule state. The transition states are
indicated in green. The harvesting was performed with transition path
sampling, making 8,400 moves in trajectory space, of which 75\% were
shooting and 25\% were shifting~\protect\cite{Bolhuis01_1}.  We find
that the plateau regime of the flux correlation function is reached
after 50-70~ps~\protect\cite{Chandler78,Bolhuis01_1} implying that the
typical commitment time for trajectories to pass over the barrier is
of the order of 0.1ns. Given this time and the fact that the figure
shows the free energy barrier separating the extended coil and compact
globule states to be about $9k_{B}T,$ the half life of the extended
chain is about 0.1~ns$ \,\times \,\exp \left( 9 \right ) \approx
10^{-5}$~s.}
\end{figure}

Figure~\ref{fig:tps.path} illustrates a 1.5~ns trajectory in which a
chain
of 12 hydrophobic units collapses from an extended coil to a compact
globule. The volume occupied by each separate unit is typical of that
for amino acids in solution.
In vacuum, this particular chain will always remain in a
coil state
since the global minimum in its intrachain potential energy function
corresponds to the fully extended conformation. Nevertheless, the
collapsed
chain is very much the stable state of the solvated chain, as is
evident
from the free energy surface shown in Fig.~\ref{fig:umbr_tps}. While
rare,
the trajectory illustrated in Fig.~\ref{fig:tps.path} is a
representative
example of the sub-ensemble of all 1.5~ns trajectories that do
exhibit the
collapse transition at temperature $T\approx 300~$K. The polymer is
initially in the extended coil state, after which a spontaneous
fluctuation
in chain conformation collapses a section of the polymer. The
collapsed
section forms a sufficiently large hydrophobic cluster that the
formation of
a vapor bubble is induced. Eventually, this vapor bubble grows and
drives
all the units of the polymer together.

Visual inspections of this and similar dynamical trajectories suggest
that
the mechanism of the collapse transition arises from an interplay
between
the size of the polymer and the formation of a vapor bubble. For this
reason, we have mapped out the free-energy landscape in terms of
variables
manifesting the polymer size and the bubble size. The first of these
variables, $R_{g}^{2}$, is the squared radius of gyration of the
polymer.
The other, $U$, is the volume of the largest vapor bubble, in units
of $l^{3}
$. In order to identify the bubbles in the system, we use the
criterion that
any two vapor cells (i.e., cells for which $n_{i}=0$) belong to the
same
bubble whenever they are nearest neighbor cells.

The contour plot for the room temperature free energy landscape,
Fig.~\ref
{fig:umbr_tps}, shows that the path of lowest free energy from the
coil to
the collapsed globule is one where, initially, the radius of gyration
decreases, while the size of the largest bubble is still essentially
zero.
In this regime, the solvent still wets the polymer (i.e., $n_{i}$ is
mostly $%
1$ for cells occupied by the solute), and the free energy hardly
changes.
When the radius of gyration becomes small enough, however, a bubble
starts
to grow. Here, the free energy increases sharply by about $9k_{B}T$
where it reaches a saddle point at $%
(U,R_{g}^{2})=(98,23.5l^{2})$. Beyond that saddle point, the bubble
grows
spontaneously to a size that eventually envelopes the fully collapsed
globule. The height of the barrier is only weakly dependent upon chain
stiffness since it is due primarily to solvent reorganization.

\newpage

A close examination of the free-energy landscape reveals a small
barrier of
height $2k_{B}T$ at $(U,R_{g}^{2})=(6.5,70l^{2})$. This feature, not
visible
on the scale of the graph plotted in Fig.~\ref{fig:umbr_tps},
separates the
coil from a more compact metastable intermediate. The presence of
these two
states arises from a competition between the entropy of the chain,
which
favors the coil state, and weak depletion forces, which favor a more
compact
state. Weak depletion forces are caused by the reduction in the
volume from
which the solutes exclude the solvent, when solutes come together
while
still wet. The attraction between two small hydrophobic objects
  predominantly arises from this effect ~\cite{Hummer96,Pratt77,Pangali79},
though
its full description (with the characteristic oscillations in
potentials of
mean force) requires a more accurate evaluation of $\Delta \mu _{{\rm
ex}%
}\left( v_{i}\right) $ than Eq.~\ref{eq:deltamu}. Whether evaluated
to high
accuracy or approximately, this driving force is relatively small,
and it is
not responsible for spontaneous assembly of hydrophobic units~\cite
{Watanabe86}. The free energy difference is $30k_{B}T$ between the
coil and
the fully collapsed globule, whereas it is only a few $k_{B}T$
between the
coil and the intermediate state. The strong driving force for the
collapse
of the polymer comes from the emptying of cells (i.e., solvent
cavitation or
drying) and thus the demixing of the hydrophobic units and water. The
resulting stabilization follows from the fact that the solvent is
close to
liquid-vapor equilibrium so that the large length scale interfacial
free
energy of the cavity is far lower than the small length scale
entropic cost
of maintaining a wet state, $\Sigma _{i}$ $\Delta \mu _{{\rm
ex}}\left(
v_{i}\right) $. Thus, nucleating the collapsed hydrophobic chain
involves
the same physical effect that is responsible for the crossover
phenomenon
discussed above in reference to Fig.~\ref{fig:bdmuoA_R.IS}.

To judge the extent to which $R_{g}^{2}$ and $U$ describe the
pertinent
dynamics of the collapse, and are thus good reaction coordinates, we
have
performed extensive transition path sampling of an ensemble of
trajectories,
each of length 150~ps. Points taken from these transition
trajectories are
shown in juxtaposition with the free energy surface in
Fig.\ref{fig:umbr_tps}%
. While ten times shorter than the one illustrated in
Fig.\ref{fig:tps.path}%
, the 150~ps trajectories are sufficiently long to capture the
mechanism of
the collapse because 150~ps is significantly longer than the time for
a
trajectory starting at the dynamical bottleneck to commit to a basin
of
attraction. Figure \ref{fig:umbr_tps} shows that the flow through the
bottleneck in this system is primarily due to the solvent, and the
solvent
in this transition state regime moves on the time scale of
picoseconds. To
identify the bottleneck (i.e., transition state surface), we have
located
the configurations on each trajectory from which newly initiated
trajectories have equal probability of ending in the coil or globule
states.
These points in configuration space are members of the transition
state
ensemble~\cite{Bolhuis01_1,Du98}. We project them onto Fig.~\ref
{fig:umbr_tps} where it is seen that points in the transition state
ensemble
are indeed close to the saddle point in the free energy surface.
Thus, $U$
and $R_{g}^{2}$ are the predominant reaction coordinates for this
system.
The transition state ensemble is slightly tilted in the $\left(
U,R_{g}^{2}\right) $-plane, showing that the larger the polymer, as
measured
by its radius of gyration, the larger the size of the critical vapor
bubble.
However, the scatter of the transition state ensemble from a line in
that
plane is notable, indicating that at least one other variable in
addition to
$U$ and $R_{g}^{2}$ plays a quantitative role in the reaction
coordinate.

The transition paths shown in Fig.~\ref{fig:umbr_tps} differ from the
lowest
free energy path, and the differences are generally larger than
$k_{B}T.$
This behavior shows that the polymer and the solvent move on
significantly
different time scales when passing through and moving beyond the
dynamical
bottleneck. When a vapor bubble is nucleated, the polymer does not
respond
on the time scale at which the bubble is formed. In fact, by taking an
artificially large value for $M_{{\rm s}}$ (3,600) to accelerate the
dynamics of the chain relative to that of the solvent by a factor of
100,
the projected dynamical paths closely follow the lowest free energy
path.

\section*{Discussion}

Trajectories of the hydrophobic collapse are generally parallel to
$U$ as they pass over the transition state. This finding demonstrates
unambiguously that the dynamics of collapse of a hydrophobic polymer 
in water is
dominated by the dynamics of water, specifically the collective emptying of
regions of space near a nucleating cluster of hydrophobic species. 
Further, both
the solvent and chain remain out of equilibrium throughout the collapse
transition. Real hydrophobic molecules have some affinity to water,
and vice versa, on account of ubiquitous van der Waals interactions.
The ideal hydrophobic chain considered here, however, has no such
property as it is composed simply of hard spheres.  Nevertheless, we
expect our conclusions to remain valid for more realistic models of
hydrophobic species because the underlying mechanism of nucleation is
largely unaffected by van der Waals interactions \cite{Huang02}.
Our findings would seem also pertinent to the mechanism of
biological assembly, such as protein folding, but to demonstrate so
with
simulation will require an analogous simulation study of a
protein-like
chain. In the past, a statistically meaningful study of such folding
with
explicit account of solvent dynamics seemed impractical. The current
analysis, however, suggests that such studies will become feasible by
using
a statistical field model like that employed herein.

Appropriate models for hydrophobic collapse must account for both the
physics of small length scale fluctuations and the physics of phase
equilibria and interfaces. This accounting is difficult to accomplish
without an explicit treatment of the solvent, as we do here. Often,
solvation is treated with implicit solvent models in which {\it all}
solvent
degrees of freedom are averaged out in some approximate fashion,
assuming
all these degrees of freedom relax quickly. Some of these models
mimic the
effects of the solvent through estimates of the pair potential of
mean force
between solute units. This approach can be correct for describing
small
length scale effects, and it therefore can be correct in situations
where
the solute remains wet. But for collections of hydrophobic solutes
large
enough to induce drying or cavitation, an implicit model would need to
describe correlations of essentially all orders, and not simply pair
correlations. For this reason, a recent study of hydrophobic effects
in
protein folding requires further analysis~\cite{Onuchic02}. Other
implicit
solvent models are based upon estimates of exposed surface area~\cite
{Eisenberg86,Vallone98}. This approach can be valid for large enough
solutes
provided the time scales of interest are longer than nucleation times
(10$%
^{-5}~$s in the system studied here). But applying it to small
unassembled
solutes is incorrect because the free energy in this regime scales
with
excluded volume and not surface area.

The connection we stress between hydrophobic collapse and phase
equilibria
provides a simple explanation for both cold denaturation and pressure
denaturation of proteins. Namely, the tendency for drying and its
concomitant stabilization of hydrophobic clustering decreases with
decreasing proximity of liquid-vapor equilibrium. Thus, the lowering
temperature and the increasing of pressure destabilize hydrophobic
collapse
because both these actions move ambient water away from its phase
equilibrium with vapor. The proximity of phase equilibria is also of
relevance to free energy barriers for hydrophobic collapse, like that
illustrated in Fig.~\ref{fig:umbr_tps}. These barriers will increase,
for
example, with increasing external pressure. Thus, the most direct
measure of
the importance of hydrophobic collapse in protein folding may
therefore come
from studying the kinetic effects of changing pressure.

\section*{Acknowledgments}

This work has been supported in its initial stages by the National
Science
Foundation (Grant No. 9508336 and 0078458) and in its final stages by
the
Director, Office of Science, Office of Basic Energy Sciences, of the
U.S.
Department of Energy (Grant No. DE-AC03-76SF00098). We are grateful
to Aaron
Dinner, David Huang and Shura Grosberg for helpful comments on an
earlier
draft of this paper.

\end{document}